\documentclass[12pt,twoside]{article}
\usepackage{fleqn,espcrc1}

\usepackage{epsfig}

\newcommand{\Msol}{\rm{M_{\odot}}}
\newcommand{\He}{\rm{^{4}He}}
\newcommand{\Ox}{\rm{^{16}O}}
\newcommand{\Si}{\rm{^{32}Si}}
\newcommand{\Ti}{\rm{^{44}Ti}}
\newcommand{\Ni}{\rm{^{56}Ni}}

\newcommand{\ie}{{\it i.e.,}\,}

\newcommand{\AmS}{{\protect\the\textfont2
  A\kern-.1667em\lower.5ex\hbox{M}\kern-.125emS}}

\title{Non-spherical core collapse supernovae and nucleosynthesis}

\author{Konstantinos Kifonidis, Ewald M\"uller and Tomasz Plewa 
        \thanks{permanent address: Nicolaus Copernicus 
                Astronomical Center, Warsaw, Poland;
                partially supported by the KBN grant 2.P03D.014.19 } \\
        Max-Planck-Institut f\"ur Astrophysik \\
        P.O. Box 1147, D-85741 Garching, Germany
       }
       
\begin{document}

\maketitle

\begin{abstract}

Motivated by observations of supernova SN\,1987A, various authors have
simulated Rayleigh--Taylor (RT) instabilities in the envelopes of core
collapse supernovae (for a review, see \cite{Mueller98}). The
non-radial motion found in these simulations qualitatively agreed with
observations in SN\,1987A, but failed to explain the extent of mixing
of newly synthesized $^{56}$Ni quantitatively.  Here we present
results of a 2D hydrodynamic simulation which re--addresses this
failure and covers the entire evolution of the first 5 hours after
core bounce.
\end{abstract}

\section{Model description and numerical setup}

Our simulation is split into two stages.  The early evolution ($t <
1$\,s) is computed with the {\sc HERAKLES} code (Plewa \& M\"uller, in
preparation) which solves the multidimensional hydrodynamic equations
using the direct Eulerian version of the Piecewise Parabolic Method
\cite{CW84} and which incorporates the physics (equation of state,
neutrino source terms and light bulb) described in the simulations of
neutrino driven supernovae by \cite{JM96} with the following
modifications. General relativistic corrections are added to the
gravitational potential.  A 14-isotope network is implemented to
compute the explosive nucleosynthesis including the 13 $\alpha$-nuclei
from $\He$ to $\Ni$ and a tracer nucleus which is used to keep track
of the neutronization of the material.  The 2D spherical grid consists
of 400 radial zones ($3.17 \times 10^{6}\,{\rm cm} \leq r \leq
1.7\times10^9$\,{\rm cm}) and 192 angular zones ($0 \leq \theta \leq
\pi$).  The initial data is a 15\,$\Msol$ progenitor model
\cite{WPE88} which has been collapsed \cite{Bruenn93}.  The
following set of neutrino parameters is adopted (cf. \cite{JM96}):
$L_{\nu_e}^0 = 3.094 \times 10^{52}\,{\rm erg/s} , L_{\nu_x}^0 = 2.613
\times 10^{52}\,{\rm erg/s}, \Delta Y_l = 0.0963, \Delta \varepsilon =
0.0688$. The neutrino spectra and the functional form of the
luminosity decay are the same as in \cite{JM96}.  A random initial
seed perturbation is added to the velocity field with a modulus of
$10^{-3}$ of the (radial) velocity of the post--collapse model.  The
computation begins 20\,ms after core bounce and is carried up to
800\,ms when the explosion energy has saturated at $E_{\rm expl} =
1.77\times10^{51}$\,erg.

The subsequent propagation of the shock through the stellar envelope
and the growth of RT instabilities is simulated with the adaptive mesh
refinement code {\sc AMRA} (Plewa \& M\"uller, in preparation).
Neutrino physics is not included in AMRA because it does influence the
explosion dynamics only within the first few seconds.  Newtonian
self-gravity is taken into account by solving Poisson's equation in
one spatial dimension with an angular average of the density, which is
an adequate approximation once the shock has left the iron core.  The
equation of state includes contributions from photons, non-degenerate
electrons, $\rm e^+e^-$-pairs, $\rm ^1H$, and the nuclei included in
the reaction network.

The AMR calculations are started with the inner and outer boundaries
located at $r_{\rm in}=10^8$\,cm and $r_{\rm out}=4.8 \times
10^{10}$\,cm.  No further seed perturbations are added.  The maximum
resolution is equivalent to that of a uniform grid of $3072 \times
768$ zones.  The radial extent of the base grid is extented whenever
the supernova shock approaches the outer grid boundary, which attains
a maximum value of $r_{\rm out} = 3.9 \times 10^{12}$\,cm at $t=
2095\,$s.  Reflecting (outflow) boundary conditions are imposed in
angular (radial) direction.  The initial data for the AMR simulation
consists of three different parts. Interior to 17\,000 km
($1.63\,\Msol$) data from the simulations of the first stage of the
explosion, and exterior to that radius data from Bruenn's 1D
post-collapse model are used.  Matched to Bruenn's model (covering
only parts of the He core) is a new progenitor model of Woosley
(private communication; data for the 1988 model are no longer
available).

\section{Results and discussion}

The simulation of the early evolution shows, in accordance with
\cite{JM96}, that neutrino driven convection leads to the formation of
a roughly spherical but very inhomogeneous post--shock shell
containing dense $\Ni$--rich regions and low--density, deleptonized
bubbles. These inhomogeneities provide the seed for RT mixing in the
stellar envelope at the Si/O and (C+O)/He interfaces of the SN\,1987A
progenitor model within only about a minute after core bounce
\cite{Kif00}.

The stellar metal core is completely shredded only 5\,minutes after
bounce and high velocity clumps of newly synthesized elements are
observed to be ejected up to the outer edge of the He--core. While the
instability is turning the inner core of the star inside out, a dense
shell forms at the He/H interface as a result of the deceleration of
the main shock in the H--envelope.  The subsequent interaction of the
metal--enriched clumps with this dense He--shell leads to their strong
deceleration. After entering the shell the clumps reach transonic
speeds and dissipate a large fraction of their kinetic energy. As a
result the entire dense shell and the H--envelope are pervaded by
bow-shocks and strong acoustic waves from 3000 to 10\,000\,s after
core bounce.  During this interaction, which has not been reported in
any previous calculation of RT instabilities in SNe\,II progenitors,
the composition within the clumps themselves is almost entirely
homogenized. Furthermore, acting like a wall the shell shields the
H--envelope of the star from becoming enriched with freshly
synthesized elements.

According to a linear stability analysis the He/H interface should be
RT unstable. However, we do not find a strong growth of the
instability at this interface.  Since the supernova shock is almost
perfectly spherically symmetric when it emerges from the He--core the
evolution in these layers proceeds basically one-dimensional.  Only
when the metal--enriched clumps reach the inner boundary of the dense
shell behind the He/H interface and start to dissipate their energy
about 3000\,s after bounce, are perturbations from spherical symmetry
induced by the waves which are thereby excited.  However, as the
interface is only moderately unstable at this time, only small scale
variations are observed. If this result also holds for other
progenitor models, it indicates that neutrino driven convection alone
is not able to provide the perturbations which are needed to induce
strong mixing of the helium core and the H--envelope as observed in
SN\,1987\,A.

Already within the first 300\,s of the explosion elements like $\Ox$
and $\Si$ that made up the original metal core as well as the newly
synthesized $\Ni$ have been mixed almost homogeneously throughout the
inner $2.0 \Msol$, \ie throughout about the inner half of the
He--core.  The extent of mixing increases up to $3.2 \Msol$ at
10\,000\,s.  Species which are not mixed this far out in mass are
$\Ti$ and our neutronization tracer. These nuclei were synthesized in
the innermost layers of the ejecta which were located very close to
the collapsed core.

\begin{figure}[t]
\begin{center}
\epsfig{file=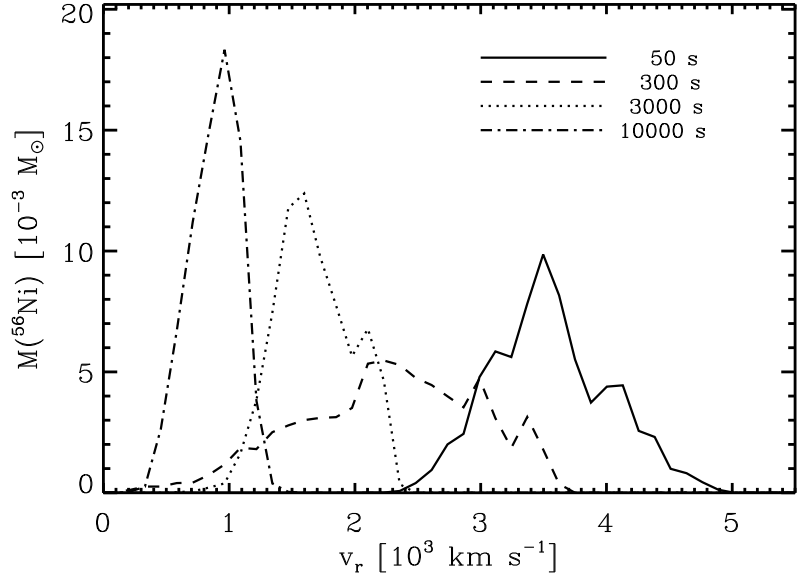, width=0.75\linewidth}
\end{center}
\vspace{-1.5cm}
\caption[]{Mass of $\Ni$ which is contained within the velocity
interval $[v_r,v_r+{\rm d}v_r]$ as a function of the radial velocity
$v_r$ at various epochs. The resolution is ${\rm d}v_r \approx
130\,{\rm km\,s^{-1}}$.}
\label{fig:nivel}
\end{figure}

The dynamics of the explosion is reflected in the distribution of the
mass of $\Ni$ in velocity space (Fig.~\ref{fig:nivel}). The most
conspicuous feature which can be seen in this plot is the bulk
deceleration of the material from velocities as large as $\sim
5000\,$km/s at a time of 50\,s after bounce to less than 1500\,km/s
after 10\,000\,s. At 50\,s the average $\Ni$ velocity is $\sim
3500$\,km/s with a spread of about $\pm 1400\,$km/s. At 10\,000\,s the
corresponding values are $\sim 900\,$km/s and $\pm 500\,$km/s.  The
maximum velocities are significantly smaller than the ones which have
been observed in SN\,1987A.  During no phase of the evolution we do
see an acceleration of material from the former metal core of the
star. This is in contrast to the results of Herant \& Benz (1992) who
report to have obtained nickel velocities comparable to those observed
in SN\,1987A provided that they premixed the $\Ni$ in their SPH
calculations throughout 75\% (in mass) of the metal core ($\sim 2
\Msol$) of their $20\,\Msol$ progenitor.

The occurrence of conditions which could give rise to the development
of an instability at the He/H interface is harmful for the propagation
of the clumps. Even if very strong perturbations are imposed upon the
dense unstable shell which forms at this interface the instability is
growing too slowly in order to shred this ``wall'' before the clumps
will reach it.  On the other hand, with a smoother density profile
between the He--core and the H--envelope, the dense shell might either
not form at all or become at least less pronounced which might help
the clumps to preserve most of their energy.  Such a smoother density
profile could be envisaged when the progenitor star of SN\,1987A was
not the result of the evolution of a single star but a merger of two
smaller stars \cite{Pod92}.  The dimensionality of our simulations may
also be a possible cause of the problem.  The fingers found in 2D
calculations are in fact axially symmetric tori which also experience
a larger drag when propagating outwards than genuinely 3D mushroom
structures.

Finally, ``missing physics'' in the explosion models itself might be
responsible for the small maximum nickel velocities that we
obtain. Effects like rotation or anisotropic neutrino emission might
have to be considered for the explosion mechanism. These may result in
an additional large scale asphericity of the shock or even in jet-like
outflows of the ejecta.  In that case it has been claimed \cite{NSS98}
that it is possible to reproduce nickel velocities in excess of
3000\,km/s. However, these simulations suffer from coarse resolution
and from rather unrealistic parameterized initial conditions.

\section{Summary}

Our calculations prove for the first time that convective
instabilities which develop during the first second of the explosion
are able to provide the seed for significant RT mixing at the Si/O and
(C+O)/He interfaces. This might offer an explanation for the mixing
observed in the explosions of SNe\,Ib/Ic \cite{Kif00}.

During the first $\sim 30$ minutes of the explosion our $\Ni$
velocities are as high as those measured in the Type\,II SN\,1987A and
the Type\,IIb SN\,1993J.  The main reason for this agreement is the
subsonic ballistic motion of the clumps relative to the mean
background flow. The concordance with observations is, however,
destroyed, when the clumps enter the outer helium core and encounter a
dense (RT unstable) shell which is left behind by the shock at the
He/H interface. The clumps are decelerated to velocities $<
2000\,$km/s because their propagation in this new environment is
transonic. Hence, the main difficulty for future simulations is how to
avoid to decelerate the initially fast $\Ni$ clumps.

\end{document}